\documentclass[journal]{IEEEtran}

\usepackage[mode=buildmissing]{standalone}

\ifCLASSINFOpdf
\else
\fi

\usepackage{amsmath}
\usepackage{amssymb}

\usepackage[acronym]{glossaries}
\usepackage{bm}
\usepackage{siunitx}
\usepackage{upgreek}
\usepackage{xfrac}
\usepackage{kitcolors}
\usepackage{tikz}
\usepackage{pgfplots}
\usepackage{calc}
\usepackage[final=true,breaklinks=true]{hyperref}
\usepackage[normalem]{ulem}
\makeatletter
\let\MYcaption\@makecaption
\makeatother

\usepackage[font=footnotesize]{subcaption}

\makeatletter
\let\@makecaption\MYcaption
\makeatother

\usepackage{mleftright}
\usepackage{booktabs}
\usepackage[noadjust]{cite}

\newif\ifautorefabbr
\autorefabbrtrue

\pgfplotsset{
  discard if/.style 2 args={
    x filter/.code={
      \edef\tempa{\thisrow{#1}}
      \edef\tempb{#2}
      \ifx\tempa\tempb
      
      \fi
    }
  },
  discard if not/.style 2 args={
    x filter/.code={
      \edef\tempa{\thisrow{#1}}
      \edef\tempb{#2}
      \ifx\tempa\tempb
      \else
      
      \fi
    }
  }
}

\let\Re\relax
\let\Im\relax
\DeclareMathOperator{\Re}{Re}
\DeclareMathOperator{\Im}{Im}
\DeclareMathOperator{\e}{e}
\let\j\relax
\DeclareMathOperator{\j}{j}

\definecolor{cb-1}{HTML}{4477AA}
\definecolor{cb-2}{HTML}{EE6677}
\definecolor{cb-3}{HTML}{228833}
\definecolor{cb-4}{HTML}{CCBB44}
\definecolor{cb-5}{HTML}{66CCEE}
\definecolor{cb-6}{HTML}{AA3377}
\definecolor{cb-7}{HTML}{BBBBBB}

\pgfplotscreateplotcyclelist{cb list}{
    cb-1,cb-2,cb-3,cb-4,cb-5,cb-6,cb-7
}

\newacronym{bps}{BPS}{blind phase search}
\newacronym{rpn}{RPN}{residual phase noise}
\newacronym{awgn}{AWGN}{additive white Gaussian noise}
\newacronym{gs}{GS}{geometrical shaping}
\newacronym{qam}{QAM}{quadrature amplitude modulation}
\newacronym{snr}{SNR}{signal to noise ratio}
\newacronym{bce}{BCE}{binary cross entropy}
\newacronym{bmi}{BMI}{bitwise mutual information}
\newacronym{gcs}{GCS}{geometric constellation shaping}
\newacronym{gmi}{GMI}{generalized mutual information}
\newacronym{mi}{MI}{mutual information}
\newacronym{e2e}{E2E}{end-to-end}
\newacronym{cpe}{CPE}{carrier phase estimation}
\newacronym{llr}{LLR}{log likelikood ratio}
\newacronym{nn}{NN}{neural network}
\newacronym{tx}{Tx}{transmitter}
\newacronym{rx}{Rx}{receiver}
\newacronym{fec}{FEC}{forward error correction}
\newacronym{bmd}{BMD}{bit-metric decoder}
\newacronym{ff-nn}{FF-NN}{feed-forward neural network}
\newacronym{ber}{BER}{bit error rate}
\newacronym{dsp}{DSP}{digital signal processing}
\newacronym{cps}{CPS}{carrier phase synchronization}
\newacronym{ml}{ML}{machine learning}
\newacronym{air}{AIR}{achievable information rate}
\newacronym{geopcs}{GeoPCS}{joint geometric and probabilistic constellation shaping}
\newacronym{pgcs}{pGCS}{parametrized geometric constellation shaping}
\newacronym{bicm}{BICM}{bit-interleaved coded modulation}
\newacronym{pas}{PAS}{probabilistic amplitude shaping}
\newacronym{dpssfm}{DP-SSFM}{dual polarization split-step Fourier method}
\newacronym{rrc}{RRC}{root-raised-cosine}

\colorlet{KITColor1}{kit-blue100}
\colorlet{KITColor2}{kit-orange100}
\colorlet{KITColor3}{kit-purple100}
\colorlet{KITColor4}{kit-maigreen100}
\colorlet{KITColor5}{kit-cyanblue100}
\DeclareSIUnit\bit{bit}
\DeclareSIUnit\symbol{symbol}
\usetikzlibrary{fit, positioning, calc, angles, arrows.meta, bending, backgrounds}
\usepgfplotslibrary{groupplots}
\usetikzlibrary{pgfplots.groupplots}
\pgfplotsset{compat=1.17}

\usepackage{algorithm}
\usepackage{algpseudocode}

\algblockdefx{cIf}{cEndIf}
[1]{\tikzmk{A}\vspace*{1ex}\textbf{if} #1 \textbf{then}}
[2][0.95]{\tikzmk{B}\textbf{end if}\vspace*{1ex}\boxit[#1]{#2}}
\algcblock[cbox]{cIf}{Else}{cEndIf}
\algrenewcommand{\algorithmiccomment}[2][.47]{\hfill\normalsize$\triangleright$ \parbox[t]{#1\linewidth}{#2}}

\newcommand*{\tikzmk}[1]{\tikz[remember picture,overlay,] \node (#1) {};\ignorespaces}
\newcommand{\boxit}[2][.95]{\tikz[remember picture,overlay]{\node[yshift=1ex,fill=#2,opacity=.25,fit={(A)($(B)+(#1\linewidth,-2ex)$)}] {};}\ignorespaces}

\makeatletter
\DeclareRobustCommand{\rvdots}{%
  \vbox{
    \baselineskip4\p@\lineskiplimit\z@
    \kern-\p@
    \hbox{.}\hbox{.}\hbox{.}
  }}
\makeatother

\pgfplotsset{colormap/blackwhite}
\pgfplotsset{colormap/bluered}
\pgfplotsset{colormap={kit-cm}{color={KITblue} color={KITorange}}}

\newsavebox\neuralnetwork
\sbox{\neuralnetwork}{%
		\begin{tikzpicture}[
      >=stealth,
      scale=.9,
      every node/.append style={transform shape},
      remember picture,
      ]%
      \tikzset{Source1b/.style={rectangle, draw=black, thick, minimum width=0.05cm, minimum height=1.2cm, rounded corners=0.5mm}}
      \tikzset{Source3/.style={rectangle, draw, thick, minimum width=0.6cm, minimum height=0.45cm, rounded corners=0.5mm}}

      \tikzset{OnehotNode/.style={circle, thick, draw,minimum width=0.1cm}}
      \tikzset{ReLUNode/.style={circle,thick,draw,fill=black!10!white}}
      \tikzset{MZMNode/.style={circle,thick,draw,fill=black!30!white}}
   \node (serial) at (3.6,0) {};

    \def\k{0}
    \def\ki{1}
        \node [OnehotNode] (S\ki1) at ($(0,0.5)+(0,\k)$) {};
        \node [OnehotNode] (S\ki2) at ($(0,-0.5)+(0,\k)$) {};
		\node at ($(0,0)+(0,\k)$) {$\rvdots$};
		\draw [->] ($(S\ki1)+(-0.4cm,0)$) -- (S\ki1);
		\draw [->] ($(S\ki2)+(-0.4cm,0)$) -- (S\ki2);

		\node [ReLUNode] (H\ki11) at ($(1,1.5)+(0,\k)$) {};
		\node [ReLUNode] (H\ki12) at ($(1,0.5)+(0,\k)$) {};
		\node [ReLUNode] (H\ki13) at ($(1,-0.5)+(0,\k)$) {};
		\node [ReLUNode] (H\ki14) at ($(1,-1.5)+(0,\k)$) {};
		\node at ($(1,0)+(0,\k)$) [anchor=center] {$\rvdots$};

		\foreach\j in {1,2,3,4}{
		  \draw [->] (S\ki1) -- (H\ki1\j); \draw [->] (S\ki2) -- (H\ki1\j);
		};

		\node [ReLUNode] (H\ki21) at ($(2,1.5)+(0,\k)$) {};
		\node [ReLUNode] (H\ki22) at ($(2,0.5)+(0,\k)$) {};
		\node [ReLUNode] (H\ki23) at ($(2,-0.5)+(0,\k)$) {};
		\node [ReLUNode] (H\ki24) at ($(2,-1.5)+(0,\k)$) {};
		\node at ($(2,0)+(0,\k)$) [anchor=center] {$\rvdots$};

		\foreach\j in {1,2,3,4}{
		  \foreach\i in {1,2,3,4}{
		    \draw [->] (H\ki1\j) -- (H\ki2\i);
		  };
		};

		\node [MZMNode] (M\ki1) at ($(3,1)+(0,\k)$) {};
		\node [MZMNode] (M\ki2) at ($(3,0)+(0,\k)$) {};
		\node [MZMNode] (M\ki3) at ($(3,-1)+(0,\k)$) {};
		\draw [<-] ($(serial.west)+(0,\k)+(0,1)$) -- (M\ki1.east);
		\draw [<-] ($(serial.west)+(0,\k)$) -- (M\ki2.east);
		\draw [<-] ($(serial.west)+(0,\k)+(0,-1)$) -- (M\ki3.east);

		\node at ($(3,-0.5)+(0,\k)$) [anchor=center] {$\rvdots$};
		\node at ($(3,+0.5)+(0,\k)$) [anchor=center] {$\rvdots$};
		\foreach\j in {1,2,3,4}{
		  \foreach\i in {1,2,3}{
		    \draw [->] (H\ki2\j) -- (M\ki\i);
		  };
		};
\end{tikzpicture}%
}

\definecolor{cR0}{RGB}{0, 0, 255} %
\definecolor{cR1}{RGB}{140,182,60} %
\definecolor{cR2}{RGB}{162,34,35} %
\definecolor{cR3}{RGB}{35,161,224} %

\definecolor{cR12}{RGB}{223,155,27} %
\definecolor{cR13}{RGB}{0,150,130} %
\definecolor{cR23}{RGB}{163,16,124} %

\definecolor{cR123}{RGB}{167,130,46} %
\definecolor{cR03}{RGB}{255, 0, 187} %

\newcommand{\RevI}[1]{#1}
\newcommand{\RevII}[1]{#1}
\newcommand{\RevIII}[1]{#1}

\begin{document}
\title{End-to-end Optimization of Constellation Shaping for Wiener Phase Noise Channels with a Differentiable Blind Phase Search}
\author{Andrej~Rode,~Benedikt~Geiger,~\IEEEmembership{Graduate~Student~Members,~IEEE,} Shrinivas~Chimmalgi, and~Laurent~Schmalen,~\IEEEmembership{Fellow,~IEEE}%
\thanks{

Parts of this work have been presented at the Optical Fiber Communications Conference (OFC) 2022 in paper \cite{rodeGeometricConstellationShaping2022a} and at the European Conference on Optical Communication (ECOC) 2022 in paper \cite{rodeOptimizationGeometricConstellation2022b}.

A. Rode (rode@kit.edu), B. Geiger, S. Chimmalgi, and L. Schmalen are with the Communications Engineering Lab (CEL) at Karlsruhe Institute of Technology (KIT).}}

\markboth{Journal of Lightwave Technology}%
{Rode \MakeLowercase{\textit{et al.}}: End-to-end Optimization of Constellation Shaping for Wiener Phase Noise channels with a Differentiable Blind Phase Search}

\maketitle

\begin{abstract}
  As the demand for higher data throughput in coherent optical communication systems increases, we need to find ways to increase capacity in existing and future optical communication links. To address the demand for higher spectral efficiencies, we apply end-to-end optimization for joint geometric and probabilistic constellation shaping in the presence of Wiener phase noise and carrier phase estimation. Our approach follows state-of-the-art bitwise auto-encoders, which require a differentiable implementation of all operations between transmitter and receiver, including the DSP algorithms. In this work, we show how to modify the ubiquitous blind phase search (BPS) algorithm, a popular carrier phase estimation algorithm, to make it differentiable and include it in the end-to-end constellation shaping. By leveraging joint geometric
  and probabilistic constellation shaping, we are able to obtain a robust and
  pilot-free modulation scheme improving the performance of 64-ary communication
  systems by at least \num{0.1}\,bit/symbol compared to square QAM
  constellations with neural demappers and by \num{0.05}\,bit/symbol compared to previously presented approaches applying only geometric constellation shaping.
\end{abstract}


\IEEEpeerreviewmaketitle

\section{Introduction}
\IEEEPARstart{I}{n} recent years, more and more innovations, e.g., internet of
things, 6G, and video streaming, continue to increase the demand for higher data
rates. Optical fiber communication systems, in particular, have to bear the bulk
of the traffic to interconnect geographical regions to provide connectivity to
said innovations. Therefore, to keep up with the growing demand, increasing the
network capacity is necessary. Ideally, the data rate of the physical layer employed
in existing communication links shall be increased.

One way to increase the data rate is to increase the spectral efficiency of optical communication
systems by applying constellation shaping. Both probabilistic and geometric shaping
achieve a shaping gain over classical square \gls{qam} constellations.
Probabilistic shaping changes the probability of occurrence of constellation
symbols arranged in the classical square \gls{qam}, while geometric shaping
changes the position of constellation points. For the \gls{awgn} channel,
there exist good solutions for probabilistic~\cite{kschischangOptimalNonuniformSignaling1993a,bochererProbabilisticShapingForward2019} and
geometric~\cite{forneyMultidimensionalConstellationsIntroduction1989,
  forneyMultidimensionalConstellationsII1989, larssonGoldenAngleModulation2018} constellation
shaping. For channels with memory or non-linearities, closed-form analytical
solutions for neither probabilistic nor geometric shaping are currently
available and we need to resort to numerical optimization techniques. Especially
for optical fiber communication systems,  a popular
approach to optimize constellation shaping is to apply \gls{ml}~\cite{jonesDeepLearningGeometric2018, karanovEndtoendDeepLearning2018, gumusEndtoendLearningGeometrical2020, dzieciolGeometricShaping2D2021}. An \gls{e2e} approach allows us to
optimize constellations that maximize the
\gls{air}~\cite{cammererTrainableCommunicationSystems2020} in ubiquitous \gls{bicm}-like
systems with \glspl{bmd}. Classical approaches for constellation shaping usually lack efficient ways to optimize the bit labeling for such \gls{bicm} system. In more
recent works \cite{arefEndtoendLearningJoint2022,
  starkJointLearningGeometric2019, oliariHybridGeometricProbabilistic2022}, joint
geometric and probabilistic shaping optimized with the help of machine learning and in particular \glspl{nn} was successfully
investigated.

Another way to increase the effective net data rate is to reduce the number of transmitted
pilot symbols. A portion of pilots
are inserted to aid \gls{cpe}~\cite{mazurOverheadoptimizationPilotbasedDigital2019}; the application
of blind \gls{cpe} reduces the need for pilot symbols. \RevIII{Reducing or fully
  eliminating the need for pilot symbols for \gls{cpe} will be a benefit for
  future systems.} A popular and widely used algorithm for
blind \gls{cpe} in optical communication systems is the
\gls{bps} algorithm. Its popularity stems from the fact
that it can be implemented in a parallel and pipelined
fashion~\cite{pfauHardwareefficientCoherentDigital2009}, enabling \gls{cpe} for
high symbol rates. This is a big advantage over decision-directed \gls{cpe}
algorithms with feedback, which are popular in wireless communication systems
because of their lower computational complexity. Due to the required feedback connection, a feed-forward, pipelined implementation is not straightforward.

Combining constellation shaping with auto-encoders and blind \gls{cpe} is
challenging, since the \gls{e2e} optimization of constellation shaping with
auto-encoders requires channel models and \gls{dsp} algorithms that are differentiable.
Differentiability is required to enable gradient descent based
optimization using the back-propagation algorithm. Other approaches try to avoid the requirement of differentiable
channels by introducing surrogate channel models~\cite{jovanovicEndtoendLearningConstellation2021a} or by using techniques like reinforcement
learning~\cite{aitaoudiaEndtoEndLearningCommunications2018}, genetic
algorithms~\cite{yanGeneticAlgorithmAided2013} or cubature Kalman
filters~\cite{jovanovicGradientfreeTrainingAutoencoders2021}. These approaches
come with a cost, as complexity of the training increases significantly and the
rate of convergence is lower. Direct \gls{e2e} optimization through back-propagation and
gradient descent is the most robust approach, since the effects of all elements
between the sender and receiver, including \gls{dsp} algorithms, are included during the training through the loss function. In order to have
an \gls{e2e} differentiable auto-encoder channel, all operations between
encoder and decoder \glspl{nn} have to be differentiable w.r.t the trainable
parameters in the \glspl{nn}. The regular \gls{bps} algorithm includes a non-differentiable $\arg\min$
operation and hence cannot be used directly with gradient descent. Thus constellation shaping
with surrogate channels or without requiring a differentiable channel have been
studied for this particular
use-case~\cite{jovanovicEndtoendLearningConstellation2021a,
  jovanovicGradientfreeTrainingAutoencoders2021, jovanovicGeometricConstellationShaping2022a}. We however want to leverage the robustness and convergence speed of gradient descent and hence in~\cite{rodeGeometricConstellationShaping2022a} proposed a differentiable \gls{bps} algorithm. In this work, we propose to use the differentiable \gls{bps} algorithm from~\cite{rodeGeometricConstellationShaping2022a}, to optimize constellations for the Wiener phase noise channel in an \gls{e2e} manner. Besides geometric constellation shaping, we also investigate the potential of joint probabilistic and geometric constellation shaping.

The remainder of this work is organized as follows. In \autoref{sec:diff_bps}, we
explain how the non-differentiable $\arg\min$ operation can be replaced with the
differentiable \emph{softmin with temperature} operation to implement the
differentiable \gls{bps} algorithm. We introduce the system model for \gls{gcs} in
\autoref{sec:gcs_model}. The extension of the \gls{gcs} system model to
\gls{geopcs} is highlighted in \autoref{sec:geopcs_model}. In \autoref{sec:simulation_setup}, we discuss the chosen simulation parameters for the \gls{e2e} optimization and discuss results we obtain for constellation shaping with the differentiable \gls{bps} for the Wiener phase noise channel. We compare and discuss the results of our optimizations of \gls{gcs} and \gls{geopcs} in \autoref{sec:results}. In particular, we show a novel approach to jointly optimize geometric and probabilistic shaping for the Wiener phase noise channel including the \gls{bps} algorithm. We highlight differences to previous approaches optimizing the geometric shaping. We conclude the paper in \autoref{sec:conclusion}.

\section{Differentiable Blind Phase Search}\label{sec:diff_bps}
In a first step, we summarize the \gls{bps} algorithm as it was presented in
\cite{pfauHardwareefficientCoherentDigital2009} and introduce our modification
to make the \gls{bps} differentiable in a second step. The \gls{bps} algorithm
is described in \autoref{alg:bps} for a set of $M=2^m$ constellation
symbols $\mathcal{M} := \{\text{c}_1,\ldots,\text{c}_M\}, \text{c}_i \in
\mathbb{C}$ and a sequence of complex received symbols $\bm{z} =
(z_1,\ldots,z_k,\ldots)$. The \gls{bps} is parametrized by the length of the
averaging window $2N + 1$ and the number of test phases $L$, which define the granularity with which the \gls{bps} estimates and corrects phase errors.

\begin{algorithm}
      \begin{algorithmic}
        \State  Constellation $\mathcal{M},\left| \mathcal{M} \right| = 2^m$ %
        \State Received symbol $z_k \in \mathbb{C}$ %
        \State Test phases $\bm{\varphi} \gets \left( 0, \frac{1}{L}2\pi, \ldots,
          \frac{L-1}{L}2\pi \right)^{\mathsf{T}}$ %
        \For{$\ell \gets 1,\ldots,L$}
        \State  $d_{k,\ell} \gets  \min\limits_{c \in \mathcal{M}}  \left| c -
          z_k \exp\left( -\mathrm{j}\varphi_\ell \right) \right|^2 $ %
        \EndFor
        \State  $D_{k,\ell} \gets
        \sum_{\tilde{k}=k-N}^{k+N}d_{\tilde{k},\ell} \qquad\forall \ell\in\{1,\ldots, L\}$%
        \vspace{1ex}
        \If{differentiable}
        \State $\hat{\phi}_{k} \gets \bm{\varphi}^{\mathsf{T}}\mathrm{softmin}_t\left(\bm{D}_k \right)$ %
        \Else{}
        \State $\hat{\ell}_{k} \gets \arg\min\limits_{\ell=1,\ldots, L} D_{k,\ell}$ %
        \State $\hat{\phi}_{k} \gets \varphi_{\hat{\ell}_k}$
        \EndIf
        \vspace{1ex}
        \State Phase unwrapping $\tilde{\bm{\phi}} \gets \mathrm{unwrap}(\hat{\bm{\phi}})$
        \State $\hat{x}_k \gets z_k \exp\left( -\j\tilde{\phi}_k \right)$
      \end{algorithmic}
      \caption{Differentiable and Regular \acrshort{bps}}
      \label{alg:bps}
\end{algorithm}
\begin{figure}[t!]
  \centering
  \includestandalone[width=0.9\columnwidth]{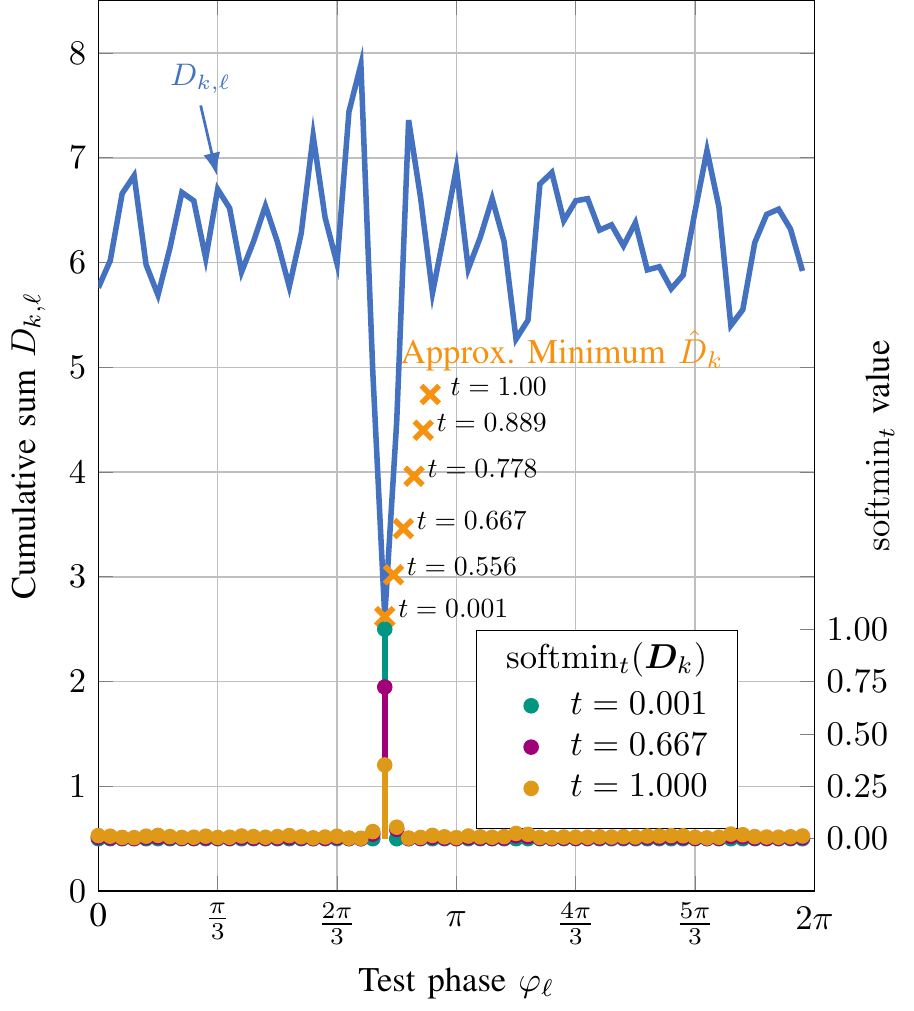}
  \caption{$\mathrm{softmin}_t(\bm{D}_{k})$ applied on \RevI{\sout{an exemplary }} the cumulative sum $D_{k,\ell}$ to approximate $\arg\min$. %
  \RevII{The approximated minimum values $\hat{D}_{k}$ marked in the plot by ``$\times$" are given by $\hat{D}_k = \bm{D}_k^{\mathsf{T}}\mathrm{softmin}_t(\bm{D}_k)$.}}
  \label{fig:softmin_t}
\end{figure}
\begin{figure*}[t!]
  \centering
  \includestandalone[scale=0.5]{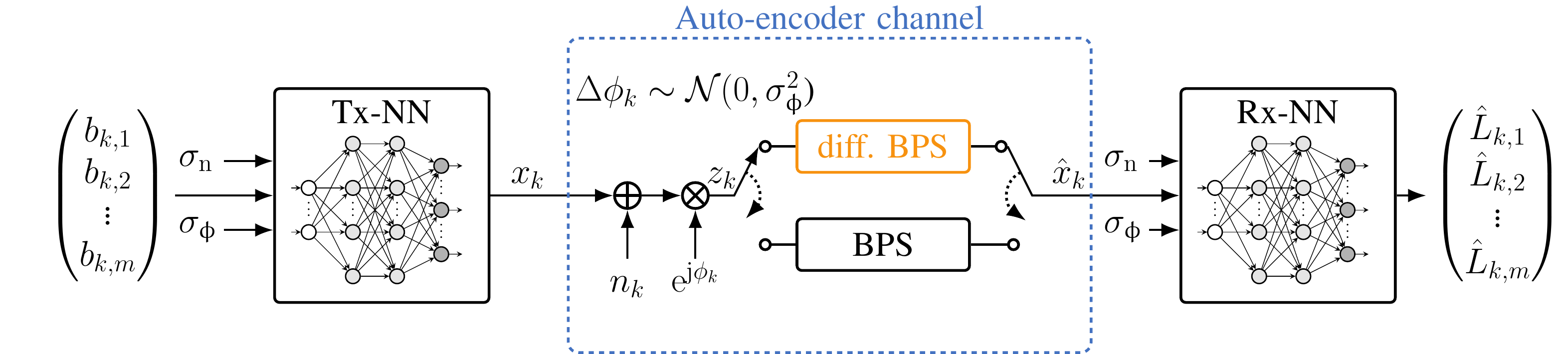}
  \caption{Bitwise auto-encoder system model for GS with differentiable BPS}
  \label{fig:system_model}
\end{figure*}

The first step in the \gls{bps} algorithm when estimating the phase error $\hat{\phi}_k$ of received symbol $z_k$ consists of finding the
minimum distance $d_{k,\ell}$ between all known transmit constellation
symbols $\text{c}_i$ and the symbol $z_k$ rotated by all test phases $\varphi_\ell$ with $\ell \in \{1, \ldots, L\}$. In an equivalent, practical, implementation, the rotated constellations can be saved and the minimum distance for all test phases can be computed in parallel. After computing the distances $d_{k,\ell}$, an averaging step is performed by calculating the cumulative sum $D_{k,\ell}$ per test phase $\varphi_\ell$ using $N$ previous and $N$ following distances. This step increases the robustness against \gls{awgn} while decreasing the resilience to higher laser linewidths. The cumulative sum introduces fringe effects at the start and the end of a \gls{bps} block, which we will need to remove during optimization to resemble the behavior of a system operating in a steady state.
After computing the cumulative sum $D_{k,\ell}$ the following operations between
differentiable \gls{bps} and regular \gls{bps} differ. With regular \gls{bps},
the test phase index $\hat{\ell}_k$ that minimizes the cumulative sum
$D_{k,\ell}$ is found using the $\arg\min$ operation and the corresponding test
phase $\hat{\phi}_k =\varphi_{\hat{\ell}_k}$ is used as a carrier phase estimate. We apply phase unwrapping to $\hat{\bm{\phi}}$ and obtain $\tilde{\bm{\phi}}$, which we use to correct the received symbols $\bm{z}$. In the
differentiable \gls{bps}, we replace the $\arg\min$ operation with a
differentiable approximation $\mathrm{softmin}_t$, which we call \emph{softmin with temperature}. In the following section, we introduce $\mathrm{softmin}_t$ in more detail.
To obtain a phase estimate $\hat{\phi}_k$, the dot product between the result of $\mathrm{softmin}_t$ and the vector of test phases $\bm{\varphi}$ has to be computed. The degree to which $\mathrm{softmin}_t$ approximates the $\arg\min$ operation is controlled with the temperature parameter~$t$.

\subsection{Softmin with Temperature}\label{subsec:softmin_temp}
We derive the $\mathrm{softmin}_t$ from the $\mathrm{softmin}$ operation. For a vector $\bm{x} =
(x_1,\ldots,x_n)^{\mathsf{T}}$, the $i$-th element of $\mathrm{softmin}_t(\bm{x})$ is given by
\begin{equation}
\mathrm{softmin}(x_i) := \left( \mathrm{softmin}(\bm{x}) \right)_i =  \frac{\exp{(-x_i)}}{\sum_{j=1}^n{\exp{(-x_j)}}}.\label{eq:softmin}
\end{equation}
 To improve the approximation of the true minimum, we scale the input vector to the softmin operation with $1/t$:
\begin{equation}
    \mathrm{softmin}_t(x_i) := \left( \mathrm{softmin}_t(\bm{x}) \right)_i =  \frac{\exp{\mleft(-\frac{x_i}{t}\mright)}}{\sum_{j=1}^n{\exp{\mleft(-\frac{x_j}{t}\mright)}}}. \label{eq:softmin_temp}
\end{equation}
Now, for $0 < t < 1$, the minimum and maximum are further separated in value than
for the unscaled input. Therefore the following $\mathrm{softmin}$ operation returns a higher weight for the minimum value.
In \autoref{fig:softmin_t}, we show the change in the approximation of
$\arg\min$ with $\mathrm{softmin}_t$ for a varying parameter $t$. Decreasing $t$
improves the approximation: For small values of $t$, most elements of the $\mathrm{softmin}_t$
output vector are zero; only a single non-zero element at a single test phase which leads to the smallest distance in the cumulative sum vector $\bm{D}_k$ persists. This corresponds closely to the $\arg\min$ operation.
\RevII{The approximated minimum value $\hat{D}_k = \bm{D}_k^{\mathsf{T}}\mathrm{softmin}_t(\bm{D}_k)$ is plotted in \autoref{fig:softmin_t} only for illustrative purposes, as we are only
interested in the phase estimate $\hat{\phi}_k$ which is given by $\hat{\phi}_k = \bm{\varphi}^{\mathsf{T}}\mathrm{softmin}_t(\bm{D}_k)$}. For $t=1$, $\mathrm{softmin}_t$ is equivalent to the
$\mathrm{softmin}$ operation and we can observe a significant offset from the true $\arg\min$.
In~\cite{maddisonConcreteDistributionContinuous2017}, the authors show how to
smoothly approximate $\arg\max$ with a similar approach applied to the softmax function. In our use case, the $\mathrm{softmin}$ operation does not approximate the $\arg\min$ sufficiently and consequently the returned phase estimate has a significant offset from the true phase unless the extra parameter $t$ is used.

\section{Geometric Constellation Shaping System Model}\label{sec:gcs_model}
Auto-encoders have been successfully used for unsupervised learning of efficient latent representations in the wider machine learning
community and are quite naturally applicable to the design of communication systems.
Namely, in an auto-encoder, information is first processed by an encoder \gls{nn} 
to get a representation in a latent space, which usually has a smaller dimension
than the input. A decoder \gls{nn} is then used to reconstruct the original
information from the information in the latent space. No labeling in the latent space is
required, instead, representations of the information can be extracted from the
latent space. We can cast this auto-encoder concept to a communication problem:
the encoder is a mapper which maps bit vectors to symbols on the
complex plane and the decoder is a demapper trying to recover the embedded
information from complex, noisy received symbols. In order to learn representations that are
efficient in the presence of channel impairments, the symbols---generated by the encoder \gls{nn} (denoted in the following by ``Tx-NN'')---are impaired by a communication channel, e.g., by \gls{awgn}, dispersion, or
multi-path propagation. These impairments are partially removed with classical \gls{dsp}
algorithms and the recovered complex symbols are processed by a decoder \gls{nn}  (denoted in the following by ``Rx-NN'')
to return a \gls{llr} vector. By optimizing encoder and decoder \glspl{nn} to
minimize the \gls{bce} loss~\cite{cammererTrainableCommunicationSystems2020}, an efficient mapping can be found by back-propagation and
gradient descent with, e.g., the Adam
algorithm~\cite{kingmaAdamMethodStochastic2015}. \RevIII{The learned encoder and
  decoder do not necessarily have to be used in a practical implementation to
  reap the benefits of this optimization. The encoder can be replaced by a
  lookup table and the bitwise decision regions can be approximated by simpler
  structure than a deep neural network. In
  \cite{rodeGeometricConstellationShaping2023}, we present a way to learn a
  decoder with comparable complexity to a Gaussian demapper.}

Our proposed system model in \autoref{fig:system_model} is built according to a
bitwise auto-encoder.
At time step $k$, a bit vector $\bm{b}_k = \left(b_{1,k}, \ldots, b_{m,k}\right)^{\mathsf{T}}$ is first
mapped to a corresponding one-hot vector $\tilde{\bm{b}}_k$ of length $2^m$ that is only
non-zero at the index that corresponds to the binary-to-integer conversion of the
bit vector. The Tx-\gls{nn} is used to generate a constellation $\mathcal{M}$ of
size $M = 2^m$. In the case of a non-parametrizable mapper, the mapper neural
network (Tx-NN) reduces to a real-valued weight matrix $\bm{W}_\text{m}$ of size
$2\times2^{m}$ and we can generate a vector of constellation symbols $\bm{c} = (\mathrm{c}_1,\ldots,\mathrm{c}_M)^{\mathsf{T}}\in
\mathbb{C}^{M}$ containing all $M$ modulation symbols of the constellation $\mathcal{M} :=
\{\mathrm{c}_1,\ldots,\mathrm{c}_{M}\}$ with
\begin{equation}
  \bm{c}^{\mathsf{T}} = \begin{pmatrix}1 & \j \end{pmatrix}
  \underbrace{\begin{pmatrix}
    W_{1,1} & \ldots & W_{1,M} \\
    W_{2,1} & \ldots & W_{2,M}
  \end{pmatrix}}_{=\bm{W}_{\mathrm{m}}}\,.
\end{equation}
The dot product between  a one-hot vector $\tilde{\bm{b}}_k$ and $\bm{c}$ is
then used to select one constellation symbol $x_k$ for transmission:
\begin{align}
    x_k &= \tilde{\bm{b}}_k^\mathsf{T} \bm{c}.
\end{align}
\Gls{awgn} and Wiener phase noise are applied to $x_k$ to simulate a
communication channel which is dominated by Gaussian noise and impairments from a non-zero laser linewidth.
The standard deviations $\sigma_\text{n} = \sqrt{N_0} =
\sqrt{\frac{E_\text{s}}{\text{SNR}} }$ and $\sigma_\upphi = \sqrt{2\pi
  \frac{\Delta f}{R}}  $ are selected so that a channel with
$\text{\gls{snr}} = \frac{E_\text{s}}{N_0}$ and linewidth $\Delta f$ at symbol
rate $R$ is simulated. The transmit symbols $x_k$ are affected by complex \gls{awgn} $n_k \sim
\mathcal{C}\mathcal{N}\mleft(0,\sigma_\text{n}^2\mright)$ and Wiener phase noise
$\phi_k = \phi_{k-1} + \Delta\phi_k$  with $\Delta \phi_k \sim
\mathcal{N}\mleft(0, \sigma_\upphi^2\mright)$. The impaired symbols $z_k$ are
then further sent through a \gls{cpe} algorithm to recover and correct the
carrier phase. In our system model in \autoref{fig:system_model}, we use either
our differentiable \gls{bps} or the regular \gls{bps}. After \gls{cpe},  we
obtain the phase compensated symbols $\hat{x}_k$. The demapper neural network
(Rx-NN) takes the complex symbols $\hat{x}_k$ and performs demapping to obtain
$m$ \glspl{llr} $\hat{\bm{L}}_k$.
\begin{figure*}[t!]
  \includestandalone[width=\textwidth]{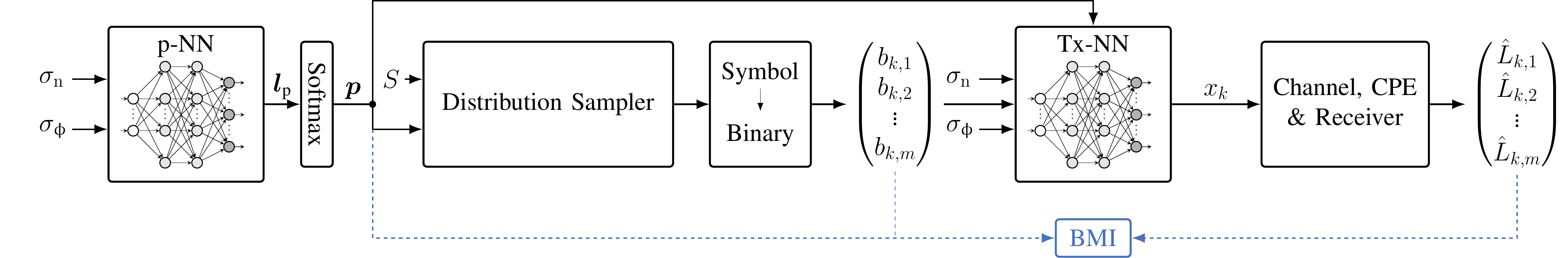}
  \caption{Parameterizable bitwise auto-encoder system model for GeoPCS}
  \label{fig:joint_system_model}
\end{figure*}

To evaluate the performance of a \gls{bicm} system we calculate the
\gls{bmi}\footnote{The \gls*{bmi} is often referred to as \gls*{gmi} in the
  optical communications community. We prefer to use the term \gls*{bmi} due to
  its easier resemblance with the operational meaning.} as the metric. For a
sequence of $P$ bit vectors $\bm{b}_k$ ($k\in\{1,\ldots, P\}$) and their corresponding
\glspl{llr} $\hat{\bm{L}}_k$, the BMI is approximated as
\begin{equation}
  \text{BMI} \approx \mathbb{H}(\mathcal{M}) -\frac{1}{P} \sum_{k=1}^{P} \sum_{i=1}^{m} \log_2 \mleft( 1 + \e^{{(-1)}^{b_{k,i}} \hat{L}_{k,i}} \mright) \label{eq:bmi_pgcs},
\end{equation}
where $\mathbb{H}(\mathcal{M})$ is the entropy in bits of the modulation symbols, calculated
from the discrete probability of occurence $\text{p}(\text{c}_i)$ for each
modulation symbol $\text{c}_i$:
\begin{equation}
  \mathbb{H}(\mathcal{M}) = -\mathbb{E}\mleft[\log_2 \text{p}(\text{c}_i)\mright].
\end{equation}
In case of a uniform distribution of the modulation symbols, \eqref{eq:bmi_pgcs} further simplifies to
\begin{equation}
    \text{BMI}\approx m - \frac{1}{P} \sum_{k=1}^{P} \sum_{i=1}^{m} \log_2 \mleft( 1 + \e^{{(-1)}^{b_{k,i}} \hat{L}_{k,i}} \mright).
    \label{eq:bmi_gcs}
\end{equation}

Subsequently, instead of a custom implementation, the \gls{bce} loss function
common to many machine learning frameworks can be used directly as a loss
function to optimize the \gls{bmi} (see~\cite{cammererTrainableCommunicationSystems2020} for details).

We initialize the Wiener phase noise process with a starting phase
$\phi_\text{start} \sim \mathcal{U}\mleft(-\pi, \pi\mright)$, where $\phi_k =
\phi_\text{start} + \sum_{k' = 0}^{k-1}\Delta \phi_{k'}$.
The random initialization of the starting phase helps to improve the robustness of the
constellation when \gls{bps} is used for \gls{cpe} without prior
compensation using pilots. This allows us to learn a constellation for a
pilot-less system which is robust to \gls{awgn} and Wiener phase noise.

\subsection{Parameterizable GCS}
Our proposed system model in \autoref{fig:system_model} contains additional
inputs $\sigma_n$ and $\sigma_\upphi$ at both
Tx-\gls{nn} and Rx-\gls{nn} to allow for the optimization of \gls{gcs} over a range
of channel parameters.  This parameterization serves multiple purposes: it
allows the investigation of changes in the constellation when channel parameters
are varied. Additionally, we obtain a mapper and demapper optimized for
each set of channel parameters within the range of training parameters. This
parameterization is roughly equivalent to training a separate set of mapper and
demapper \glspl{nn} for each individual set of channel parameters. Such individualized training would be significantly more computationally expensive and the bit labeling may be different for different channel parameters preventing us from analyzing the isolated impact of the variation in the parameters on the shaping. With this \gls{pgcs}, each symbol---which corresponds to a particular bit
vector---is moved only in a small region defined by the set of channel parameters. The \gls{pgcs} retains its general structure and changes can be
observed with respect to the change in the channel parameters.

\subsection{Trainable Differentiable BPS}
Having implemented a differentiable \gls{bps}, a natural question arises: Can the
differentiable implementation replace the regular \gls{bps} not only for the
training and optimization, but also for evaluation and implementation?
To investigate this possibility, we additionally make the temperature $t$
in the differentiable \gls{bps} trainable. To implement a
regularization of the temperature parameter $t$, we choose an unbounded trainable
parameter $t^\ast$, which we use to calculate $t=\sigma\mleft(t^\ast\mright)$.
Here, $\sigma\mleft(x\mright) = (1 + \e^{-x})^{-1}$ is the sigmoid function
and limits $t$ to the range $(0,1)$. We then use the
differentiable \gls{bps} and the optimized parameter $t^\ast$ in the
implementation instead of the regular \gls{bps}.
Since the phase estimate of the differentiable \gls{bps} is not limited to the
granularity of the test phases $\bm{\varphi}$, the
remaining residual phase noise may decrease.

\section{Joint Geometric and Probabilistic Constellation Shaping System Model}\label{sec:geopcs_model}
To incorporate the idea of learning a joint geometric and probabilistic shaping
of the constellation, we extend our system model in \autoref{fig:system_model}
with a distribution sampler at the transmitter for the training
in~\autoref{fig:joint_system_model}, as proposed in~\cite{arefEndtoendLearningJoint2022}.
This extension changes the operation of the training in terms of how bits and
symbols are generated and sent through the channel. With \gls{geopcs}, an \gls{nn} (denoted by ``p-NN'') first generates a
vector of logits $\bm{\ell}_\text{p}$  to represent the probability of
occurrence for each constellation point in the log domain. Then, a softmax
operation is applied on $\bm{\ell}_\text{p}$ to get a vector of probabilities $\bm{p}$.

For each training batch, the symbol probabilities $\bm{p}$ and the batch size $S$ are passed to a distribution sampler. To find the quantized number of symbols according to the given distribution of symbol probabilities $\bm{p}$, we apply Algorithm~2~from~\cite{bochererOptimalQuantizationDistribution2016}. We convert the symbol indices to their binary representations to get $S$ bit vectors $\bm{b}_k$, which can be passed through the neural mapper to get $S$ complex symbols $\bm{x}$. To account for the non-uniform symbol probability, we change the energy normalization operation in the neural mapper to apply $\bm{p}$ to scale the contribution of each constellation point according to the symbol probability $\text{p}(\mathrm{c}_i)$. The normalized constellation is obtained as
\begin{equation}
    \mathcal{M}_\text{norm} = \left\{{\text{c}_i}\cdot\left(\sum\nolimits_{j=1}^M \text{p}(\text{c}_j)|\text{c}_j|^2\right)^{-\frac{1}{2}}: \text{c}_i \in \mathcal{M}\right\}.
\end{equation}
The remaining operations to obtain the $S$ \glspl{llr} vectors are the same as in the case for only \gls{gs} in \autoref{fig:system_model}.
For optimization of this \gls{geopcs} system model, we use a modified loss
function based on the \gls{bmi}~in~\eqref{eq:bmi_pgcs}. This is motivated by the
fact that using the \gls{bce} does not correctly incorporate the reduction of
the entropy resulting from probabilistic shaping in~\eqref{eq:bmi_pgcs}.
The entropy is
calculated directly from $\bm{p}$ and optimization is performed with automatic
differentiation and gradient descent. We also extend the \RevI{\sout{the }}approach
presented in \cite{arefEndtoendLearningJoint2022} to use a neural network with
one hidden layer with inputs for $\sigma_\upphi$ and $\sigma_n$ to allow for
parameterization of $\bm{\ell}_\text{p}$. In the case parameterization is not required, the neural
network for $\bm{\ell}_\text{p}$ simplifies to a vector of trainable weights.

\subsection{Probability Distributions with Symmetry}
In order to learn a logits vector $\bm{\ell}_{\text{p}}$ which can be implemented jointly with
\gls{fec} in a \gls{bicm} system akin to the~\gls{pas} scheme~\cite{bochererProbabilisticShapingForward2019,schmalenProbabilisticConstellationShaping2018}, we extend the probabilistic shaping to be
parameterizable with a symmetry parameter $s$. This parameter $s$ controls the
size of the output dimension of the p-NN to be $2^{m-s}$. This limits the parameter
$s$ to a range of $[0,m-1]$ to have any probabilistic shaping. To obtain a
probability of occurence for each modulation symbol, the logits vector $\bm{\ell}_{\text{p},s}$ is
repeated $s$ times to form the final logits vector $\bm{\ell}_\text{p}$ of
length $2^{m}$. This introduces an $s$-fold symmetry in $\bm{\ell}_\text{p}$ and
in turn this results in $s$ bits of the bit vector forming the modulation
symbols that are 0 or 1 with equal probability. This property can be inferred from the
symbol-to-bit mapping and bit-to-symbol mapping, which convert the symbol index
to the binary representation. This property can then be used to combine $m-s$ information bits from a distribution matcher (in the final system) with  $s$ parity check bits from an \gls{fec} encoder to select modulation symbols corresponding to the probability distribution. For regular and  fully flexible
probabilistic shaping, the parameter $s$ is set to zero. To perform
probabilistic constellation shaping which contains one bit with equal
probability distribution the symmetry parameter has to be set to $s=1$. As highlighted in the system models in
\cite{bochererProbabilisticShapingForward2019,schmalenProbabilisticConstellationShaping2018},
a fixed number of bits with an equal probability distribution can be used to assign the approximately uniform distributed
parity check bits from~\gls{fec}.

\section{Simulation Setup}\label{sec:simulation_setup}
We perform \gls{gcs} and \gls{geopcs} with the system model shown in
\autoref{fig:system_model} and \autoref{fig:joint_system_model} with the set of
hyperparameters given in \autoref{tab:system_parameters_training}. The trainable
parameters of the \glspl{nn} are initialized randomly with the Glorot
initializer. Validation is performed with the regular \gls{bps} algorithm unless
stated otherwise. \RevI{Across epochs, the channel parameters such as SNR and
  linewidth are fixed. In each batch of every epoch, the input bits are randomly
  permuted and a new noise realization is used.} While a symbol rate of
\SI{32}{GBaud} is quite low for modern single-carrier transmission systems, it
is a realistic value for multi-carrier transmission
systems~\cite{sun800GDSPASIC2020,geigerRecord29Tb2022,welchDigitalSubcarrierMultiplexing2022a}.
\RevIII{Additionally, for low cost systems where coherent transmission is
  relevant, the symbol rate may be higher, but the resulting phase noise is
  still comparable to our parameterization due to low cost and high linewidth
  lasers \cite{shahpariCoherentAccessReview2017}.}

\begin{table}
    \centering
    \caption{System parameters for training Tx-\gls{nn} and Rx-\gls{nn}}
    \begin{tabular}{c c}
    \toprule
     Bits per Symbol $m$ & 6 \\
     Epochs    &  \num{1000} \\
     Batches per Epoch & $10, \ldots, 500$\\
     Batchsize per Epoch   & $1000, \ldots, 10000$  \\
     Optimization Algorithm & Adam~\cite{kingmaAdamMethodStochastic2015} \\
     Learning rate & $0.001$\\
     Temperature parameter $t$ & $1.0, \ldots, 0.001$ \\
     Test phases $\bm{\phi}$ & 60 \\
     Cumulative sum length $N$ & 128\\
     Symbol rate & \SI{32}{GBaud} \\
    \bottomrule
    \end{tabular}

    \label{tab:system_parameters_training}
\end{table}

We implemented our system in the PyTorch machine learning framework
\cite{NEURIPS2019_bdbca288} and our source code is accessible in \cite{rodeMLCommsPythonLibrary}.
Selection of an appropriate range of temperature values for the $\mathrm{softmin}_t$ operation during the optimization is crucial for the stability of the training. Starting with a temperature $t=1$ and then decreasing it to $t=0.001$ as the training of the \glspl{nn} progresses yielded stable optimization in our setup. We attribute this behavior to the fact that for low temperature values, only a few of the distance measures around the minimum contribute to the phase estimate. Therefore, most of the other test phases are scaled with a value close to zero; subsequently, also the contribution to the gradients is small. Therefore, a rather wide inclusion of values at the beginning of the training helps to boost the training, to avoid getting stuck in a local minimum and finally to reach a good distribution of constellation points that are optimized in later training epochs.

Training and validation for the shaped constellations has been performed with random phase initialization of the Wiener phase noise process. In the case of \gls{qam}, the Wiener phase noise has been initialized with zero and \gls{bps} only operates on the first quadrant, since otherwise the phase offset cannot be estimated correctly due to the rotational ambiguities in the \gls{qam} constellation.

\section{Results}\label{sec:results}
We performed optimization of the constellation shape with both \gls{gcs} and the
\gls{geopcs} system models. Both approaches have been trained with a fixed
channel parameterization to get a general idea about the shape and bit labeling
of the learned constellation mapper and demapper. Additionally, we also trained
with varying channel parameters as additional inputs to the \glspl{nn}.
This allows us to conduct a further investigation of the effects of \gls{cpe} on
learned communication systems and we obtain an estimate of the expected
performance gain for optimized \gls{gcs} and \gls{geopcs}.

\subsection{Fixed Channel Parametrization GCS and GeoPCS}
\begin{figure}
  \begin{subfigure}{\columnwidth/2}
    \centering
    \includestandalone[width=\textwidth]{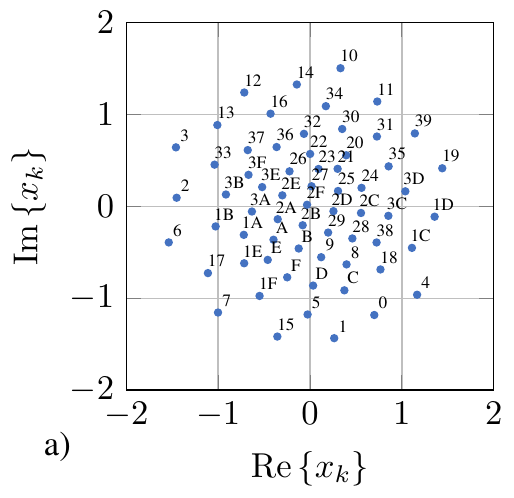}
  \end{subfigure}%
  \begin{subfigure}{\columnwidth/2}
    \centering
    \includestandalone[width=\textwidth]{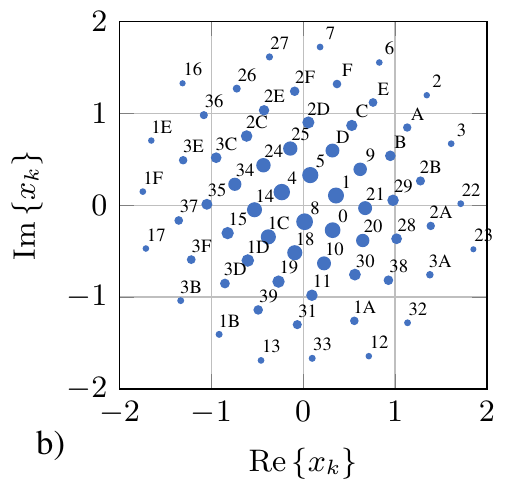}
  \end{subfigure}
    \caption{Bit-labeled constellation with a) GCS and b) GeoPCS, trained using the differentiable
      \gls{bps} algorithm at \SI{17}{dB} \gls{snr} and \SI{100}{kHz} laser
      linewidth. The size of the constellation points in b) indicates the probability
    of occurrence. Bit labels are obtained by converting the bit vectors to their
    hexadecimal representation, e.g., $(0,1,1,1,1,1) \equiv 1\text{F} $}
    \label{fig:constellation_1}
\end{figure}
In this approach, the training is performed with fixed channel parameters, which
are not passed as inputs to Tx-\gls{nn} and Rx-\gls{nn}. The optimization of the \glspl{nn} is performed at a fixed operating point, and we expect the performance on different channel parameterizations to be worse.
In \autoref{fig:constellation_1}a), we show a \gls{gs} constellation for an
$\text{SNR}=\SI{17}{dB}$ and laser linewidth $\Delta f = \SI{100}{kHz}$. A
pronounced feature is the introduced asymmetry in the constellation in the lower
left corner. Multiple constellation points are placed further separated from the other
constellation points than their counterparts in other corners. This positioning increases the robustness and supports
the operation of the \gls{bps} algorithm by resolving possible ambiguities of
constellation points. This asymmetry comes at the expense of more closely packed
constellation points at the center of the constellation and thus offers lower
robustness to \gls{awgn} compared to a constellation optimized for \gls{awgn} only.
In \autoref{fig:constellation_1}b),  we show a
\gls{geopcs} constellation with the same fixed channel parameters as the
\gls{gcs} constellation. For the human eye, the induced asymmetry is hard to
spot, but the constellation contains less rotational symmetry compared to a square \gls{qam}.

\begin{figure}[!t]
    \begin{subfigure}{\columnwidth/2}
        \includestandalone[width=\textwidth]{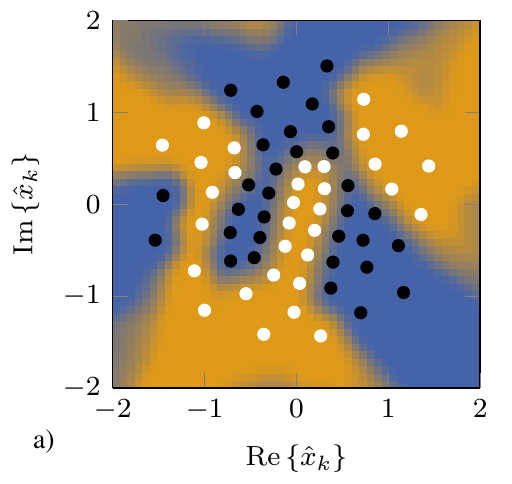}
    \end{subfigure}%
    \begin{subfigure}{\columnwidth/2}
        \includestandalone[width=\textwidth]{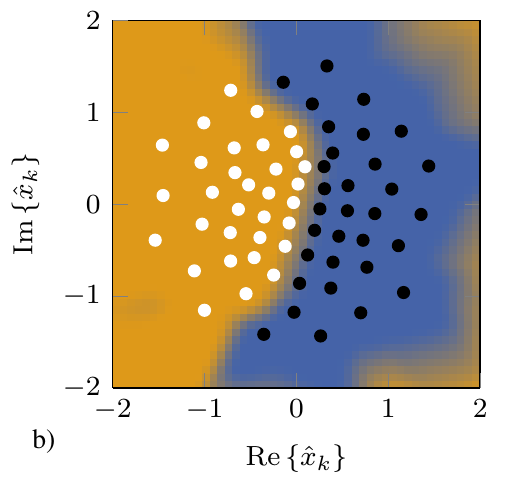}
    \end{subfigure}\\
    \begin{subfigure}{\columnwidth/2}
        \includestandalone[width=\textwidth]{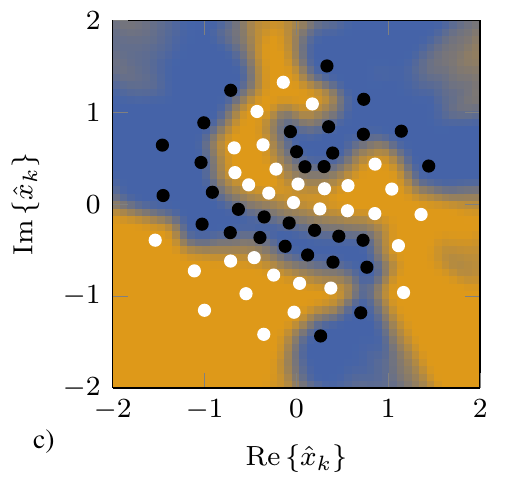}
    \end{subfigure}%
       \begin{subfigure}{\columnwidth/2}
        \includestandalone[width=\textwidth]{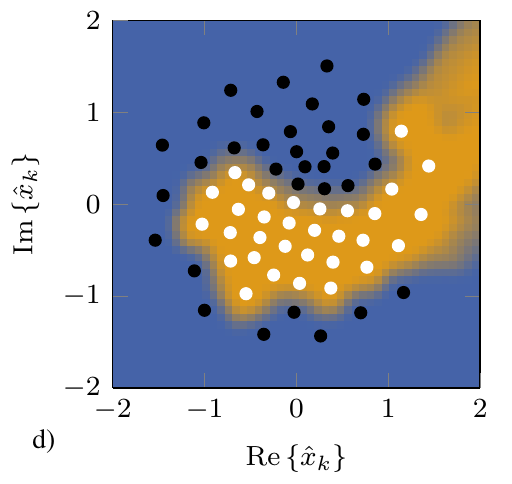}
    \end{subfigure}\\
    \begin{subfigure}{\columnwidth/2}
        \includestandalone[width=\textwidth]{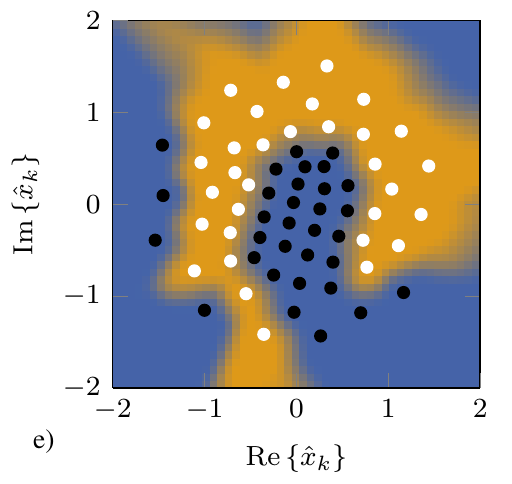}
    \end{subfigure}%
    \begin{subfigure}{\columnwidth/2}
        \includestandalone[width=\textwidth]{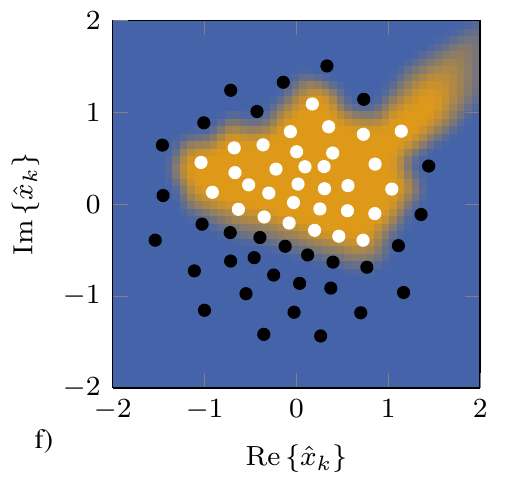}
    \end{subfigure}\\
    \begin{subfigure}{\columnwidth}
        \hfill
        \includestandalone[width=0.91\textwidth]{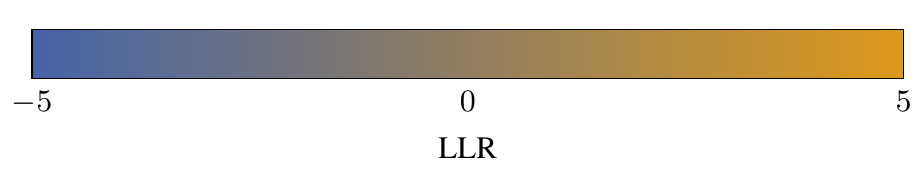}
        \hspace{0.1mm}
    \end{subfigure}
    \caption{Neural demapper decision regions for each of the bits for
      constellation in \autoref{fig:constellation_1}a). \RevII{Sub-figure a)
        corresponds to the least-significant bit and sub-figure f) to the most
        significant bit of the binary labels.}}
    \label{fig:demapper_decision_region}
\end{figure}

To investigate the performance of our \gls{e2e} system optimized
with \gls{gcs}, we take a look at the learned neural demapper and the
decision regions for individual bits as depicted in
\autoref{fig:demapper_decision_region}. For better visualization, the demapper
decision regions are shown for \glspl{llr} between $\left[-5, 5\right]$.

\begin{figure}
  \centering
  \includestandalone[width=0.9\columnwidth]{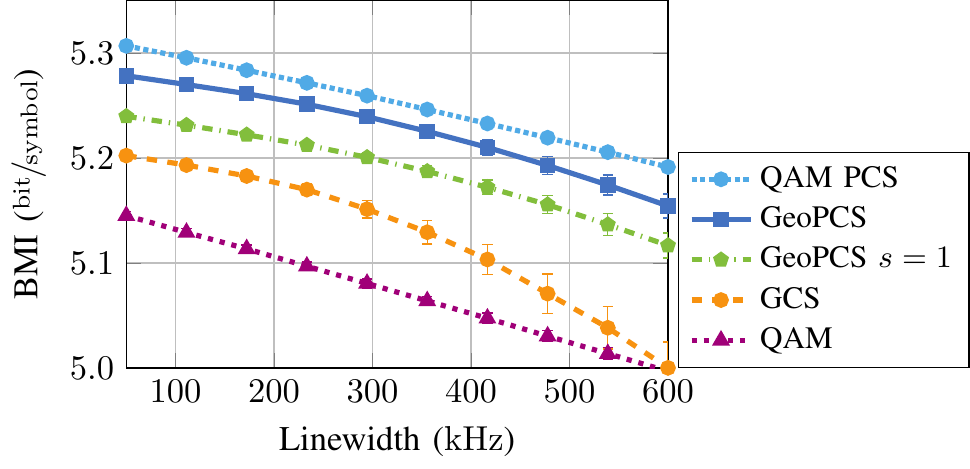}
  \caption{Performance comparison between constellations optimized at $\text{SNR}=\SI{17}{dB}$ with GCS and GeoPCS at an $\text{SNR}=\SI{17}{dB}$ and a laser linewidth $\Delta f = \SI{100}{kHz}$.}
  \label{fig:performance_pcs}
\end{figure}

With the chosen neural demapper architecture, our system is able to accurately
partition the constellation into two sets of the same size for each bit. The
decision regions shown in \autoref{fig:demapper_decision_region} are an extension
of the 1D \gls{llr} plots shown in \cite{shentalMachineLLRningLearning2019} to
the full 2D region. A 1D graph would not be enough to characterize the decision
regions of the neural demapper, since $L_{k,i}$ depends on both $\Re\{\hat{x}_k\}$ and
$\Im\{\hat{x}_k\}$.

Finally, we compare the attainable performance
of the different approaches in \autoref{fig:performance_pcs}. For
the systems trained on a fixed set of channel parameters, the \gls{geopcs} system
has the best performance, outperforming our previously
presented work on \gls{gcs}~\cite{rodeGeometricConstellationShaping2022a} by
\SI{0.05}{bit/symbol} at the channel parameters both systems were trained
on and by even more in the region of higher laser linewidths. The \gls{geopcs}
constellation with introduced symmetry also outperforms the \gls{gcs}
constellation. All shaped constellations outperform a square \gls{qam}
constellation paired with a neural demapper, which was trained on the same
channel parameters as the shaped constellation systems. For a fair comparison,
the validation of the square \gls{qam} was not initialized with a random start
phase, since \gls{qam} is rotationally symmetric and the \gls{bps} algorithm
wouldn't be able to disambiguate carrier phases if the phase difference equals a
multiple of $\pi/2$; this means that phase slips (sometimes referred to as cycle slips) do not play a role in our evaluation.
\RevI{To investigate the performance of probabilistically shaped QAM, we perform
  an \gls{e2e} optimization to learn the parameter $\lambda$ of the
  Maxwell-Boltzmann distribution for the fixed channel parameters. We apply a
  similar approach as for \gls{geopcs}, but instead of the probability of
  occurence for each symbol, the \gls{nn} returns the parameter $\lambda$, which
  is then used with the Maxwell-Boltzmann distribution and known constellation
  symbols to obtain the probability distribution. In
  \autoref{fig:performance_pcs}, we report the results of this validation as
  ``QAM PCS''. For the fixed channel parameterization we see a significant gain
  compared to square QAM and a similar performance as our proposed schemes.
  This difference in performance is most likely caused by Gray
  labeling in the square QAM and ``QAM PCS'' which is not achieved by the
  trained geometric constellations, but rather an almost Gray labeling is learned.}

\subsection{GCS with Varying Channel Parametrization}
\begin{figure}[!t]
\centering
  \includestandalone[width=\columnwidth]{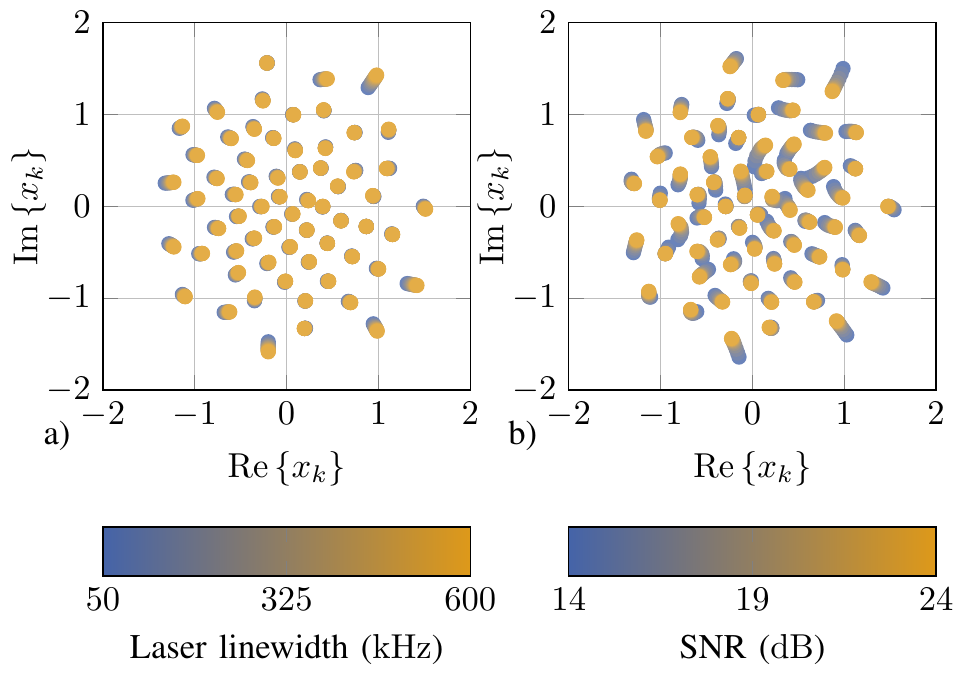}
  \caption{pGCS constellation for $M=64$ at a) fixed $\text{SNR} = \SI{18}{dB}$ and varying laser linewidth, and b) fixed laser linewidth $\Delta f = \SI{100}{kHz}$ and varying SNR}
  \label{fig:constellation_var_group}
\end{figure}

To optimize \gls{gcs} for a range of varying channel parameterizations, we train
the \gls{pgcs} system on channel parameterizations drawn from
$\mathcal{U}(\sigma_{\text{n},\min}, \sigma_{\text{n},\max})$ and
$\mathcal{U}(\sigma_{\upphi,\min},\sigma_{\upphi,\max})$. The parameters are
also provided to the \glspl{nn}, such that the system is optimized for each set
of provided channel parameters and learns how to change the constellation to
obtain a better constellation and demapper for the given condition. In
\autoref{fig:constellation_var_group}, we show the obtained transmit constellation for training the system on a channel with \gls{snr} between \SI{14}{dB} and \SI{24}{dB} and the laser linewidth $\Delta f$ between \SI{50}{kHz} and \SI{600}{kHz}.
The constellation is depicted for a fixed
$\text{SNR} = \SI{18}{dB}$ and varying
laser linewidth in a color change in \autoref{fig:constellation_var_group}a).
For a high laser linewidth, the constellation
exhibits a few points at the outside of the constellation at the top right and
bottom right corners, which move outwards compared to the constellation optimized for a
lower laser linewidth. The change in the transmit constellation diagram is small
and in stark contrast to the change in the transmit constellation diagram for
varying \gls{snr} depicted in \autoref{fig:constellation_var_group}b). In this
diagram, the laser linewidth $\Delta f = \SI{100}{kHz}$ is fixed and the color
indicates the \gls{snr}. For low \gls{snr}, the same constellation points as for
varying laser linewidth are moved outwards compared to the transmit
constellation for high \gls{snr}. Additionally, for low \gls{snr}, more points at
the center of the constellation are moved closer together to allow for a separation of
other constellation points at the outer edge of the constellation. This
``sacrifices'' the information contained in at least one bit carried by the
constellation and results in a more Gaussian-like shaping of the complete transmit
constellation diagram.

\begin{figure}[!t]
\centering
\includestandalone[width=\columnwidth]{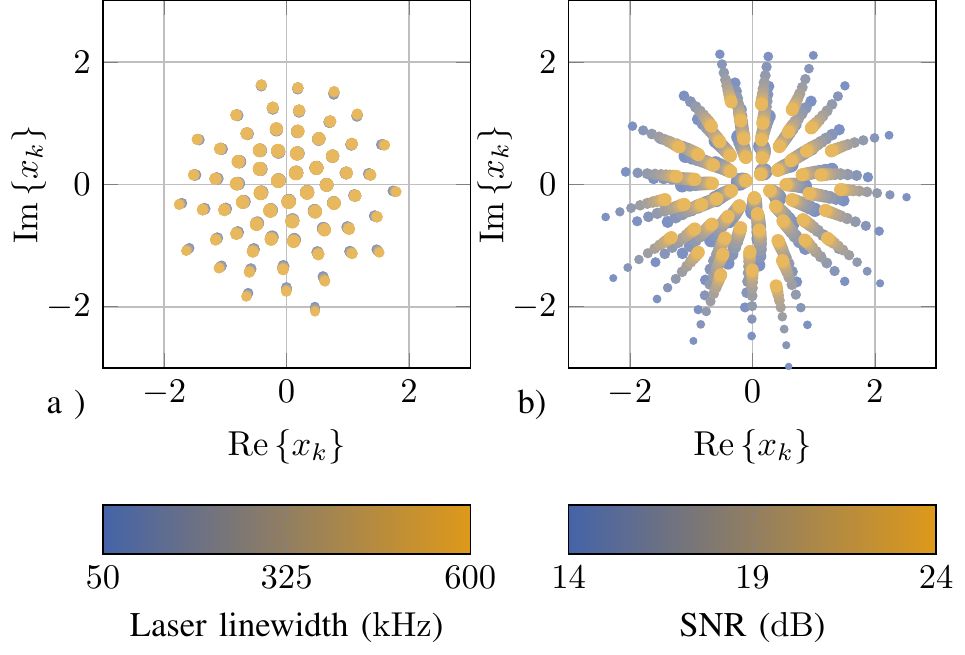}
    \caption{pGeoPCS constellation for $M=64$ at a) fixed $\text{SNR} = \SI{18}{dB}$ and varying laser linewidth, and b) fixed laser linewidth $\Delta f = \SI{100}{kHz}$ and varying SNR}
    \label{fig:constellation_var_pcs}
\end{figure}

\begin{figure}
\centering
\includestandalone[width=\columnwidth]{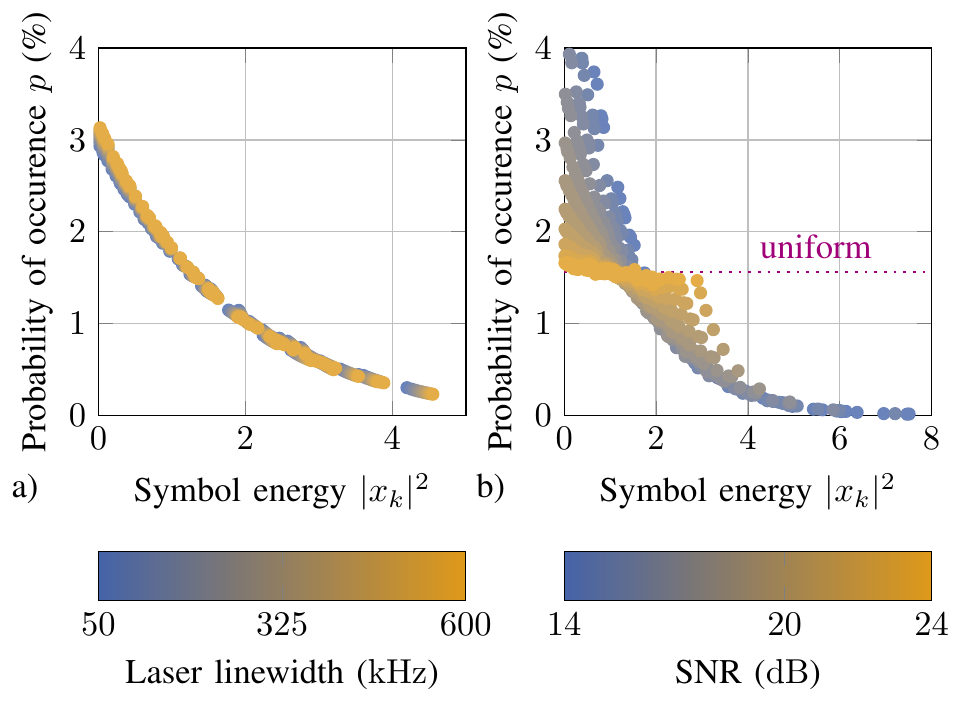}
\caption{Learned PCS for a) varying laser linewidth at SNR = \SI{18}{dB}, and b) varying SNR at laser linewidth $\Delta f = \SI{100}{kHz}$}
\label{fig:varying_pcs}
\end{figure}

The results of training the parameterized \gls{geopcs} (pGeoPCS) are depicted in
\autoref{fig:constellation_var_pcs} and \autoref{fig:varying_pcs}. The
transmit constellation is shown in~\autoref{fig:constellation_var_pcs}a) for
varying laser linewidth and shown in~\autoref{fig:constellation_var_pcs}b) for
varying \gls{snr}. The change in the probability of
occurrence  for the transmit symbols over the symbol energy is shown in
\autoref{fig:varying_pcs}a) for varying laser linewidth and for varying
\gls{snr}  in \autoref{fig:varying_pcs}b). Contrary to the pGCS system, a varying
laser linewidth does not change any features of the constellation,
neither the geometric constellation  shape in
\autoref{fig:constellation_var_pcs}a) nor the probabilistic constellation shape
in \autoref{fig:varying_pcs}a). This result is somewhat surprising, as this
means that the learned transmit constellation is only changed for varying
\gls{snr} and is independent of the laser linewidth. For a change in \gls{snr}, a behaviour similar to that of the \gls{awgn} channel is observed~\cite{arefEndtoendLearningJoint2022}. The probabilistic shaping approaches the shape of the well-known Maxwell-Boltzmann distribution.

\begin{figure}
    \centering
    \includestandalone[width=\columnwidth]{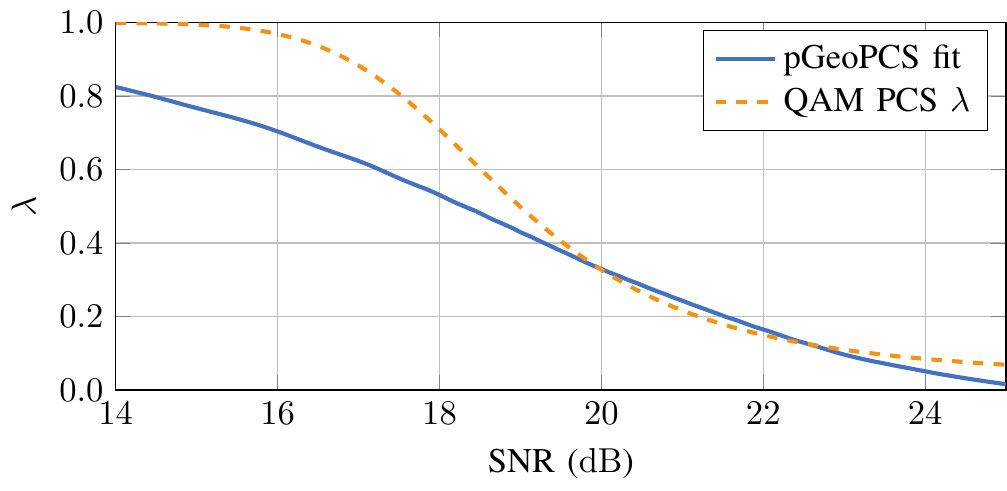}
    \caption{Numerical fit of learned PCS to parameter $\lambda$ of the
      Maxwell-Boltzmann distribution over SNR}
    \label{fig:pcs_mb_fit}
\end{figure}

\begin{figure}[!t]
  \centering
  \includestandalone[width=\columnwidth]{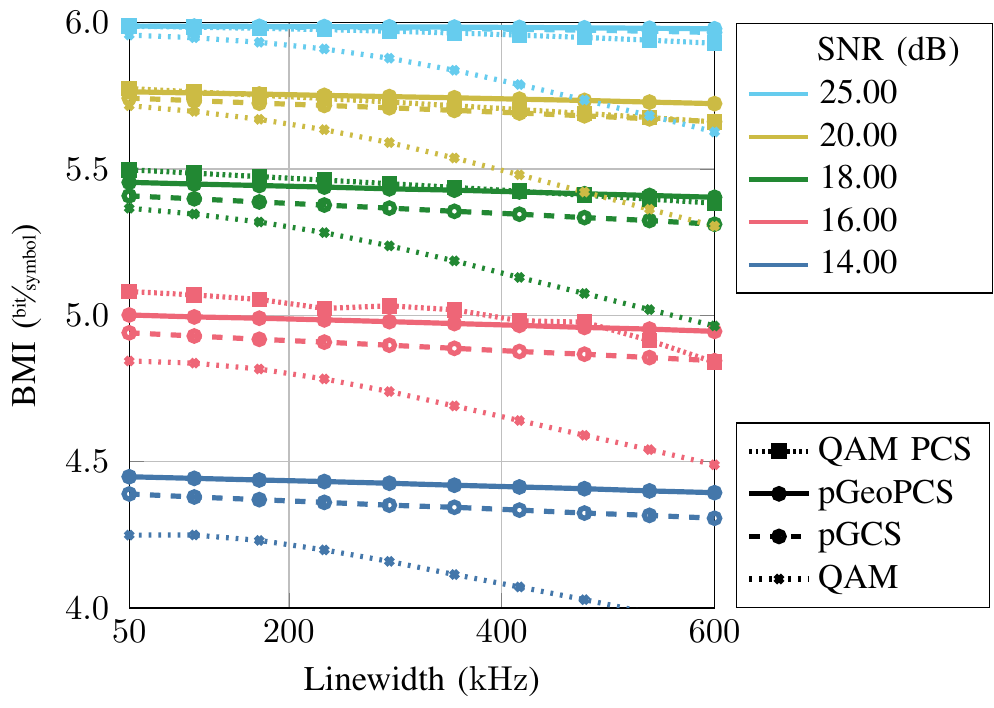}
  \caption{Performance comparison between constellations optimized with GCS and GeoPCS with varying channel parameters.}
  \label{fig:performance_pgeopcs}
\end{figure}

In \autoref{fig:pcs_mb_fit}, we depict the value $\lambda$ for the normalized Maxwell-Boltzmann probability mass function
\begin{equation}
    \text{p}_{\text{MB}}(\mathrm{c}_i) = \frac{\e^{-\lambda |\mathrm{c}_i|^2}}{\sum_{v=1}^{M} \e^{-\lambda |\mathrm{c}_v|^2}},
\end{equation}
which has the closest numerical fit to the learned symbol distribution.
The fit to the Maxwell-Boltzmann distribution
is very accurate with a maximum observed Kullback-Leibler divergence of
\num{0.0002}. \RevI{In addition, we also display the learned parameter $\lambda$
  of the parameterized ``QAM PCS'' constellation. In our training, we limited
  the $\lambda$ to the range of $[0,1]$ by applying a $\mathrm{sigmoid}$
  function at the output of the neural network. In the resulting plot, a similar
shape of the learned $\lambda$ parameter over a changing \gls{snr} can be
observed. The $\lambda$ parameter learned by the probabilistically shaped
\gls{qam} differs from the fitted $\lambda$ parameter from the pGeoPCS
constellation especially for low \gls{snr}. For a higher $\lambda$, a stronger
probabilistic shaping is performed. This difference may indicate a
reason for the performance difference seen in \autoref{fig:performance_pgeopcs}}.

In \autoref{fig:performance_pgeopcs}, we compare the performance of our different
approaches to constellation shaping with varying channel parameters. The
performance is shown in \gls{bmi} over a varying laser linewidth. Colors indicate \gls{snr} and different lines and markers indicate the
system. The \glspl{nn} of all depicted systems used $\sigma_\text{n}$ and
$\sigma_\upphi$ as input to optimize the mapper and demapper for the channel
parameters used for validation. In this comparison, applying \gls{geopcs}
provides a consistent performance gain over \gls{gcs}. But it is also clearly
visible that for higher \gls{snr}, this shaping gain disappears. \RevI{In comparison,
the results of ``QAM PCS'', which is a parameterized version of the
 probabilistically shaped square QAM (using a Maxwell-Boltzmann distribution), show stability issues for
low \gls{snr}. For the lowest value of \SI{14}{dB}, the results are not shown
since the performance was impacted by many cycle slips. In the higher
\gls{snr} regime, a similar picture as to the fixed training point can be seen:
for low laser linewidths, ``QAM PCS'' outperforms the pGeoPCS scheme, where for
higher laser linewidths pGeoPCS is more robust and shows a better performance. We
conclude from these results that the probabilistically shaped square QAM has
advantages in the low laser linewidth regime if the shaping parameter $\lambda$
is chosen correctly. For high laser linewidths, our proposed schemes employing
geometric shaping show better robustness. We have to emphasize that the square
QAM constellations have a better starting condition, their phase noise
realizations are initialized with zero phase offset, therefore mandating the use
of regular pilot symbols to retain this phase reference.}%

\begin{figure}[!t]
  \centering
  \includestandalone[width=\columnwidth]{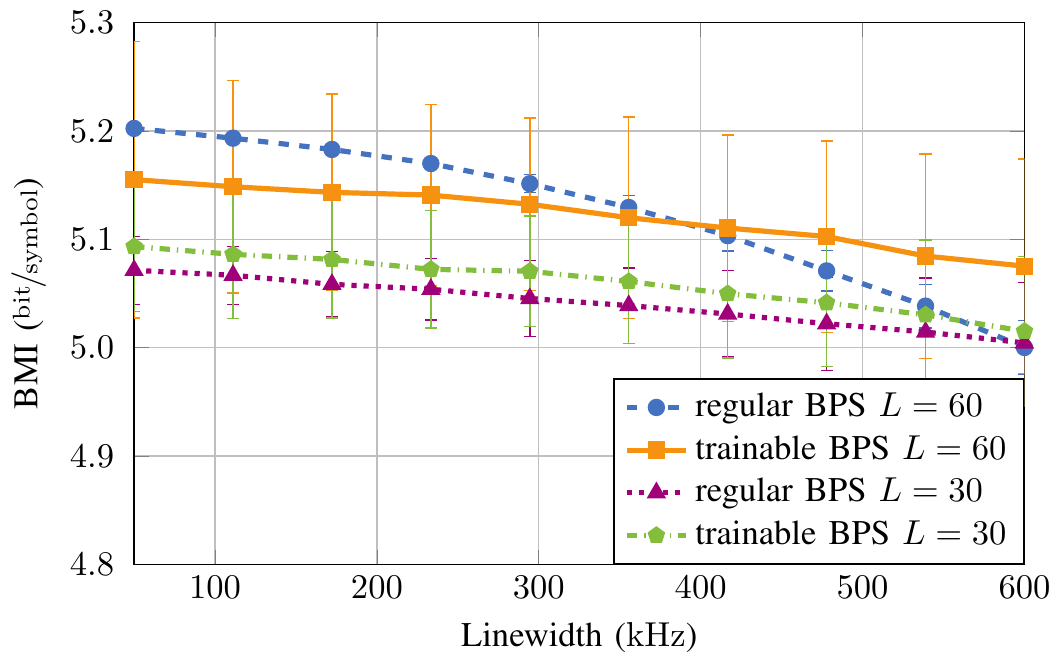}
  \caption{Performance comparison between trainable differentiable BPS and
    regular BPS trained on a channel with $\text{SNR} = \SI{17}{dB}$ and laser
    linewidth $\Delta f = \SI{100}{kHz}$.}
  \label{fig:performance_trainable_bps}
\end{figure}

\subsection{Trainable Differentiable BPS}

For the trainable \gls{bps}, our assumption is that a lower number of test phases
could be used for the final system since the differentiable \gls{bps} can
interpolate between the test phases. Therefore, we train and evaluate \gls{gcs}
systems with and without a trainable differentiable \gls{bps} for a number of
test phases $L = 30$ and $L = 60$. All systems are trained on a fixed
$\text{SNR} = \SI{17}{dB}$ and a fixed laser linewidth $\Delta f = \SI{100}{kHz}$.
The performance comparison between the system
    with trainable differentiable \gls{bps} and the system with regular \gls{bps} is
depicted in \autoref{fig:performance_trainable_bps}. For $L=60$, the regular
\gls{bps} has a consistently better performance compared to the trainable
\gls{bps}. \RevI{We attribute this to the phase discontinuity at $0$ and $2\pi$.
  If the actual phase error $\phi_k$ is close to the discontinuity, we obtain
  similar values for the cumulative sum $D_{k,\ell}$ for the test phases
  $\varphi_1 = 0$ and $\varphi_L = \frac{L-1}{L}2\pi$ in the \gls{bps} algorithm. In that case, the $\mathrm{softmin}_t$ will return high values for both these test phases even when the temperature $t$ is very low. Subsequent multiplication
and summation with the test phase vector $\bm{\varphi}$ returns a phase error estimate $\hat{\phi}_k$
around $\pi$ which is significantly different from the actual phase error leading to incorrect phase correction and consequently large errors in the \glspl{llr}. This issue does not occur with an $\arg\min$, since the issue of phase error estimates of consecutive symbols oscillating between $0$ and $2\pi$ is corrected by the phase unwrapping step after
obtaining the phase error estimates.}

The incorrect phase corrections in some cases leads to higher performance variation of the trainable \gls{bps}
between runs, which is indicated by the error bars. For $L=30$ the trainable
\gls{bps} has a slightly better performance. In this case, the ability to perform
interpolation with the $\mathrm{softmin}_t$ operation between test phases might reduce residual
phase noise compared to the regular \gls{bps}, but high performance variations
between runs are still observed. We therefore suggest further investigation into
increasing the stability and robustness of the differentiable \gls{bps} to avoid
problems at phase errors close to the discontinuity.

\subsection{\RevIII{Validation with Dual Polarization Split-Step Fourier Method Simulation}}

\RevIII{
In order to verify our results reported in~\autoref{fig:performance_pgeopcs} on
a channel impaired by chromatic dispersion, we additionally perform validation
with an implementation of a \gls{dpssfm}. This method simulates the effects
described by the coupled non-linear Schrödinger equation. We do not include
other channel variations, like polarization mode dispersion~(PMD) or  random
polarization rotation. This would entail investigating joint optimization of
constellation shaping and blind equalization
methods~\cite{lauingerBlindEqualizationChannel2022}. Similarly to a geometrically shaped
constellation with prominent asymmetric features, a constellation which is
shaped geometrically to assist a blind equalizer may exhibit other asymmetric
features. The study of interaction between the optimization of geometric
constellation shaping and equalization is an interesting topic, but is outside the scope
of this work. Applying the \gls{dpssfm} introduces some inter-symbol interference and
the results of this validation may give an indication if increased tolerance for phase noise come
at a cost of higher sensitivity to inter-symbol interference.}

\RevIII{
First, we apply a \gls{rrc} pulse shaping and upsampling to the simulation rate, to the transmit symbols $x_k$.  Then, \gls{awgn} and Wiener phase noise are added to the transmit samples to simulate transmitter impairments. The same Wiener phase noise realization is used on both polarizations. The impaired transmit samples are then sent through the \gls{dpssfm} and a different Wiener phase noise is applied on the receiver side. Afterwards we downsample the received signal and apply digital chromatic dispersion compensation. The matched filter follows the dispersion compensation and we apply the \gls{bps} algorithm separately for both polarizations. In this validation we use the regular \gls{bps}. In the case of square \gls{qam}, we apply a genie-aided phase-slip compensation on every symbol to remove any phase ambiguity by amultiple of $\pi/2$. We use the trained neural receivers to obtain \glspl{llr} and compute the performance in terms of \gls{bmi}.}

\begin{table}
  \centering
  \caption{\RevIII{Parameters for the DP SSFM}}
  \RevIII{%
    \begin{tabular}{c c}
      \toprule
      Transmitter SNR & \SI{24}{\decibel} \\
      RRC rolloff & 0.1 \\
      RRC filter length & 1000 \\
      RRC oversampling & \SI{2}{samples \per symbol} \\
      Simulation oversampling & \SI{16}{samples \per symbol} \\
      Baud rate & \SI{32}{\giga Baud} \\
      Optical carrier wavelength & \SI{1550}{\nano\metre}\\
      Span length & \SI{80}{\km} \\
      Number of spans & \num{4} \\
      SSFM steps per span & \num{80} \\
      Attenuation $\alpha_\mathrm{dB}$ & \SI[per-mode=reciprocal]{0.2}{\decibel \per \km} \\
      Dispersion $D$ & \SI[per-mode=reciprocal]{17}{\pico\second \per \nm \per \km} \\
      Kerr parameter $\gamma$ & \SI[per-mode=reciprocal]{1.2}{\per\watt \per \km} \\
      EDFA noise figure & \SI{5}{\decibel} \\
      \bottomrule
    \end{tabular}}
  \label{tab:parameters_dpssfm}
\end{table}

\RevIII{
We selected system parameters reported in \autoref{tab:parameters_dpssfm}. To
find the best launch power we performed a sweep across the launch power for each
constellation shaping method. For this validation, we use the models trained in
a parameterizable fashion for the channel with \gls{awgn}, Wiener phase noise
and the \gls{bps} algorithm as presented in \autoref{sec:geopcs_model} and
\autoref{sec:gcs_model}, and use them in the validation without any re-training.
One challenge is correctly selecting the input parameters $\sigma_\upphi$ and
$\sigma_\mathrm{n}$ for the \glspl{nn}. Since we apply i.i.d. Wiener phase noise
at the transmitter and receiver, we assume the resulting phase noise coming from
the Wiener phase processes can be described with $\sigma_{\upphi,\mathrm{tot}} =
\sqrt{2}\sigma_\upphi$ and we use $\sigma_{\upphi,\mathrm{tot}}$ as input
parameter for the \glspl{nn}. For \gls{awgn}, we have multiple sources to
account for, first we apply \gls{awgn} at the transmitter, but also the ASE
noise contribution of the EDFAs is modeled as \gls{awgn}. Additionally, the
noise contribution of the non-linearity and interaction between non-linearity
and chromatic dispersion is not accounted for in the parameterization of the \glspl{nn}. Therefore we perform a sweep of the $\sigma_\mathrm{n}$ parameter, which is only used as input to the \glspl{nn}, between \SI{17}{dB} and \SI{24}{dB} to find the best operation point for the parameterizable mappers and demappers.}

\begin{figure}
  \centering
  \includestandalone[width=0.9\columnwidth]{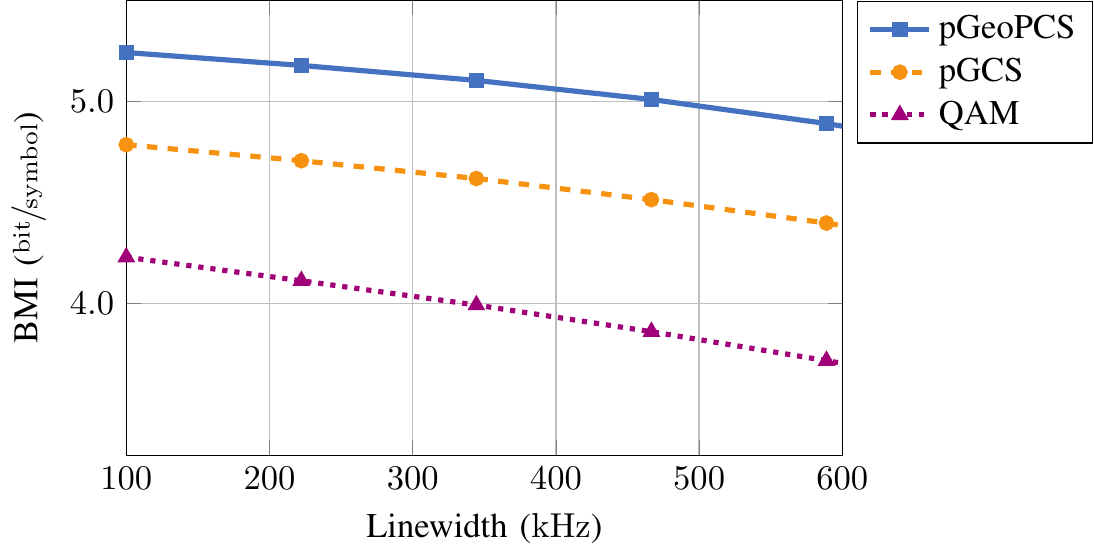}
  \caption{\RevIII{Performance comparison between square QAM, pGCS, and pGeoPCS constellations on a DP-SSFM channel. The shown laser linewidth is the sum of the laser linewidth at the transmitter and receiver, corresponding to $\sigma_{\upphi,\mathrm{tot}}$}}
  \label{fig:performance_ssfm}
\end{figure}

\RevIII{
In \autoref{fig:performance_ssfm}, we show the performance of the different constellations in terms of their \gls{bmi} over a varying laser linewidth. The launch power has been optimized between \SI{0}{dBm} and \SI{3}{dBm} in steps of \SI{1}{dBm} and for all constellations, a launch power of \SI{2}{dBm} has shown the best performance. As for the $\sigma_\mathrm{n}$ input parameter for the \glspl{nn}, the best performance for all constellations has been observed for a parameter corresponding to an $\mathrm{SNR} \approx \SI{18}{dB}$. Validation with the \gls{dpssfm} model confirms our observations made on the linear channel model in \autoref{fig:performance_pcs} and in \autoref{fig:performance_pgeopcs}. The parametrizable \gls{geopcs} constellation outperforms both \gls{pgcs} and \gls{qam} constellations. We computed the \gls{e2e} \gls{snr} in our simulations to be between \SI{19}{dB} and \SI{20}{dB}. Comparing the results in \autoref{fig:performance_ssfm} with results obtained on the linear channel, we observe a significant impact by additional non-linear impairments and residual inter-symbol interference.}

\section{Conclusion}\label{sec:conclusion}
We presented and analyzed bitwise auto-encoders for joint optimization of
geometric and probabilistic shaping for Wiener phase noise channels with carrier
phase estimation. \RevII{We used a differentiable carrier phase estimation (CPE) algorithm which allows for efficient end-to-end optimization using gradient-descent}. With the proposed system, optimization of GCS leads to a
constellation robust to phase noise and supports the operation of the CPE.
A further extension to joint geometric and probabilistic constellation shaping
provides additional shaping gain compared to GCS. In this work, we showed that joint
geometric and probabilistic shaping can be applied to communication systems
impaired by AWGN and Wiener phase noise with BPS as the carrier phase estimation algorithm. With a parameterizable mapper, demapper and probabilistic shaper,
we are able to outperform square QAM by more than \SI{0.1}{bit/symbol}
for low SNR values. \RevII{Comparison with gradient-free methods and a study of their computational complexity and convergence speed is planned future work.}

\section*{Acknowledgment}
This work has received funding from the European Research Council (ERC)
under the European Union's Horizon 2020 research and innovation programme (grant agreement No. 101001899). The authors acknowledge support by the state of Baden-Württemberg through bwHPC.

\ifCLASSOPTIONcaptionsoff
  \newpage
\fi

\bibliographystyle{IEEEtran}
\bibliography{IEEEabrv,references}

\end{document}